\documentclass[letter]{aa} 

\usepackage{graphicx}
\usepackage{txfonts}
\begin{document}

   \title{Circumnuclear dense gas disk 
   fuelling the active galactic nucleus 
   in the nearby radio galaxy NGC~4261}


   \author{
   	Satoko Sawada-Satoh\inst{1}\fnmsep\inst{2}, 
       Seiji Kameno\inst{3}\fnmsep\inst{4}, 
          \and
       Sascha Trippe\inst{5}
          }

   \institute{
   		The Research Institute for Time Studies, Yamaguchi University, 
            1677-1 Yoshida, Yamaguchi, Yamaguchi 753-8511, Japan \\
              \email{swdsth@gmail.com}
         \and
         	Graduate School of Science, Osaka Metropolitan University, 
             1-1 Gakuen-cho, Naka-ku, Sakai, Osaka 599-8531, Japan 
         \and
              Joint ALMA Observatory, 
              Alonso de C\'{o}rdova 3107 Vitacura, Santiago 763-0355, Chile 
         \and
         	NAOJ Chile Observatory, 
		Alonso de C\'{o}rdova 3788, Oficina 61B, Vitacura, Santiago, Chile
         \and
             Department of Physics and Astronomy, Seoul National University, 
             1 Gwanak-ro, Gwanak-gu, Seoul 08826, Korea
             }


 
  \abstract
 {
The cold molecular gas in the circumnuclear disk (CND) of radio galaxies  
provides critical information 
for understanding the mass accretion onto active galactic nuclei. 
We present the first detection and maps of 
HCN $J$=1--0 and HCO$^+$ $J$=1--0 emission lines 
from the circumnuclear region of a nearby radio galaxy, NGC~4261, 
using the Northern Extended Millimeter Array.  
Both molecular lines are detected 
at a radial velocity of  
$\pm700$ km~s$^{-1}$ relative to 
the systemic velocity of the galaxy, 
and they arise from a CND 
with an outer radius of ~100 pc. 
The velocity fields of HCN and HCO$^+$ 
are fitted with a Keplerian disk rotation. 
The enclosed mass is 
$(1.6\pm0.1)\times10^{9}$ $M_{\odot}$, 
assuming a disk inclination angle of $64^{\circ}$. 
The continuum image at 80 GHz reveals a weak two-sided jet structure 
extending over 5 kpc along the east--west direction 
and a bright core at the centre.  
The continuum spectrum between 80 and 230 GHz shows a spectral index 
of $-0.34\pm0.02$, 
which suggests optically thin synchrotron radiation. 
The dense gas mass associated with the CND is calculated 
to be $6.03\times10^7$ $M_{\odot}$. 
It supports a positive correlation 
between the dense gas mass in the CND  
and the accretion rate onto the 
supermassive black hole, 
though there are uncertainties 
in the parameters of the correlation. 
 }

   \keywords{
   	ISM: molecules ---
   	galaxies: active --- 
	galaxies: individual (NGC~4261, 3C~270) --- 
	galaxies: ISM --- 
	galaxies: jets --- 
	galaxies: nuclei ---
	radio lines: galaxies
            }

 \titlerunning{Circumnuclear dense gas disk in NGC~4261}
 \authorrunning{S. Sawada-Satoh et al.}

   \maketitle
%
%

\section{Introduction}\label{sec:intro}

It is widely accepted that an 
active galactic nucleus (AGN) is powered by 
mass accretion onto a supermassive black hole (SMBH)
in the centre of the host galaxy.
The gravitational energy of accreting matter is converted 
into radiation and/or jets. 
Radio galaxies (RGs) are radio-loud AGNs characterised 
by powerful synchrotron radiation 
driven by relativistic jets on scales approximately 
in the range 
10--100 kiloparsec.

The interstellar medium (ISM) in the centre of the RGs 
can play a key role in fuelling the SMBH.  
Several research groups have suggested that 
the different roles of hot and cool ISM accretion 
can be related to a different mode of accretion 
in RGs, 
which can lead to different radio-loud AGN classifications  
\citep{hardcastle07,buttiglione10,best12}. 
Certain CO observations of RGs support the hypothesis 
that RGs are fed by cold gas that probes  
the circumnuclear disks 
\citep[CNDs;][]{prandoni10,maccagni18,ruffa19}. 
The CNDs of RGs 
can serve as a reservoir of fuel for their SMBHs. 
Thus, 
by determining the molecular gas structure and kinematics of CNDs, 
important clues can be obtained 
regarding mass accretion in RGs. 
High angular resolution imaging of the molecular gas 
within the SMBH sphere of influence ($r_g$)
can also be a powerful tool for accurately measuring  
SMBH masses \citep[e.g.][]{davis13}.

However, the distribution of low-$J$ CO lines appears to extend 
to the edge of CNDs, 
and the detection of strong CO emission 
from within $r_g$ seems to be rare 
for CNDs in early-type galaxies \citep[ETGs;][]{davis18,boizelle19,north19}.  
Alternative lines may trace the molecular gas distribution 
within $r_g$ better,  
thereby enabling more accurate measurements of SMBH masses. 
At present, emission lines other than CO have largely been overlooked, 
and dense gas tracers 
such as HCN and HCO$^+$ lines are expected to trace possibly 
farther into $r_g$ than the optically thick low-$J$ CO lines. 
Moreover, recent interferometric observations of 
the dense gas emission-line tracer HCN 
towards Seyfert galaxies (SGs) 
have indicated   
that the accretion onto SMBHs is triggered 
by star formation and supernovae originating from within CNDs 
\citep{izumi16}.
To date,  star formation activities in CNDs 
have been investigated in a limited number of RGs,  
such as 
NGC~5128 \citep{espada19}, 
NGC~1052 \citep{kameno20}, 
and NGC~1275 \citep{nagai21}.
These investigations of RGs 
have been primarily conducted 
using the distribution of CO as a molecular gas mass tracer. 
Dense gas tracers can potentially 
serve as better probes 
for examining star formation activity in CNDs 
because star formation is closely related to dense gas.

NGC~4261 (3C~270) is a nearby Fanaroff--Riley I RG
with a symmetric, kiloparsec-scale two-sided jet 
\citep{birkinshaw85}. 
Its AGN is classified as a type 2 
low-ionisation nuclear emission-line 
region (LINER) galaxy \citep{jaffe96, ho97} 
 with a low Eddington ratio,  
 $L_{\rm bol}/L_{\rm Edd}$, ranging 
 from $10^{-5.11}$ to $10^{-4.54}$ 
 \citep{hernandez13, inayoshi20}
 and  
 a low X-ray luminosity of $L_{\rm 2-10 keV} = 10^{41.51}$
 erg s$^{-1}$ \citep{hernandez13}. 
This galaxy is known to have a nuclear disk of dust and gas 
with a radius of a few hundred parsecs    
lying orthogonal to the jet, 
as revealed by \textit{Hubble} Space Telescope (HST) 
observations 
\citep[][]{jaffe93,jaffe96}.

At radio frequencies, 
H~$\textsc{i}$ absorption 
has been detected at the systemic velocity of the galaxy 
($V_{\rm sys}$) towards the core 
via the Very Large Array \citep[VLA;][]{jaffe94}, 
and it has been confirmed 
at a projected distance of approximately 2.5 pc 
from the core 
with the European 
Very Long Baseline Interferometry (VLBI) 
Network \citep{vanlangevelde00}. 
The H~$\textsc{i}$ absorbing gas is interpreted to be  
in the inner part of the disk of dust and gas 
found in the HST image, 
and it obscures the core and innermost counter jet. 
Subsequent VLBI observations at multiple frequencies 
revealed the presence of parsec-scale ionised absorbing gas, 
which was likely at the inner parsec-scale radii of the HST disk 
\citep[][]{jones97,jones00,jones01,haga15}.  
Molecular lines were observed towards the centre  
of NGC~4261 for emission 
(CO $J$=2--1 and $J$=3--2; \citealt{boizelle21}) 
and absorption (CO $J$=1--0; \citealt{jaffe94}).

In this Letter, we report the first detection and 
the interferometric emission-line maps of 
HCN $J$=1--0 and HCO$^{+}$ $J$=1--0 transitions in NGC~4261, 
which trace a 100-parsec rotating CND 
perpendicular to the kiloparsec-scale radio jet. 
We adopt a luminosity distance ($D_{\rm L}$) of 31.7 Mpc and 
a $V_{\rm sys}$ of 2212 km s$^{-1}$ 
\citep[e.g.][]{babyk19,cappellari11}.
Hence, 1 arcsecond corresponds to 151 pc for the galaxy.

\section{Observations and data reduction} \label{sec:obs}

Observations were conducted on February 15, 2019, 
with the NOrthern Extended Millimeter Array (NOEMA) 
of the Institut de radioastronomie millim\'{e}trique (IRAM) 
in A configuration with ten antennas. 
The phase centre was set to the position of NGC~4261 at 
RA(J2000)=$12^{\rm h}19^{\rm m}23^{\rm s}.220$ and 
Dec(J2000)=05$^{\circ}$49$^{\prime}$30$^{\prime\prime}$.775.
The full width at half maximum of the NOEMA primary beam was 
$55^{\prime\prime}$ at 88 GHz.
The projected baseline lengths ranged from 22 to 734 m 
over the course of the observations. 
The local oscillator frequency was set to 82.0 GHz, 
with frequencies ranging from 70.398 to 78.115 GHz in the lower sideband
and from 85.886 to 93.603 GHz in the upper sideband.
The PolyFix correlator was configured with a frequency resolution of 2~MHz.
The nearby source 3C~273 was observed 
as both the bandpass and gain calibrators.
The absolute flux calibration was performed using MWC 349 (1.07 Jy) 
and LkH$\alpha$101 (0.22 Jy).
The absolute flux calibration uncertainty for NOEMA is
less than $10\%$ at Band 1 ($\lambda$3 mm)
\footnote{IRAM NOEMA Data Reduction CookBook, 
https://www.iram.fr/IRAMFR/GILDAS/doc/html/pdbi-cookbook-html/pdbi-cookbook.html}. 

The raw visibility data were first converted into 
the Flexible Image Transport System (FITS) format data 
through the GILDAS software \citep{pety05}.
Then, calibration and imaging were performed 
by using the NRAO Astronomical Image Processing System (AIPS) package 
\citep{greisen90}. 
We applied uniform weighting to the images 
to obtain a higher spatial resolution of $<1^{\prime\prime}$.
To create the continuum map, 
all line-free spectral windows with frequency ranges 
of 70.398--74.459, 74.461--78.115, and 89.948--93.603 GHz 
were combined, resulting in a centre frequency of 80 GHz. 
After the continuum emission was subtracted in the \textit{u-v} plane, 
channel maps of lines were made every 20~MHz, 
corresponding to a velocity resolution of 68 km~s$^{-1}$. 

\section{Results}
\subsection{Continuum emission}

   \begin{figure}[ht]
   \begin{center}
   \resizebox{\hsize}{!}{\includegraphics[bb=0 0 512 272]{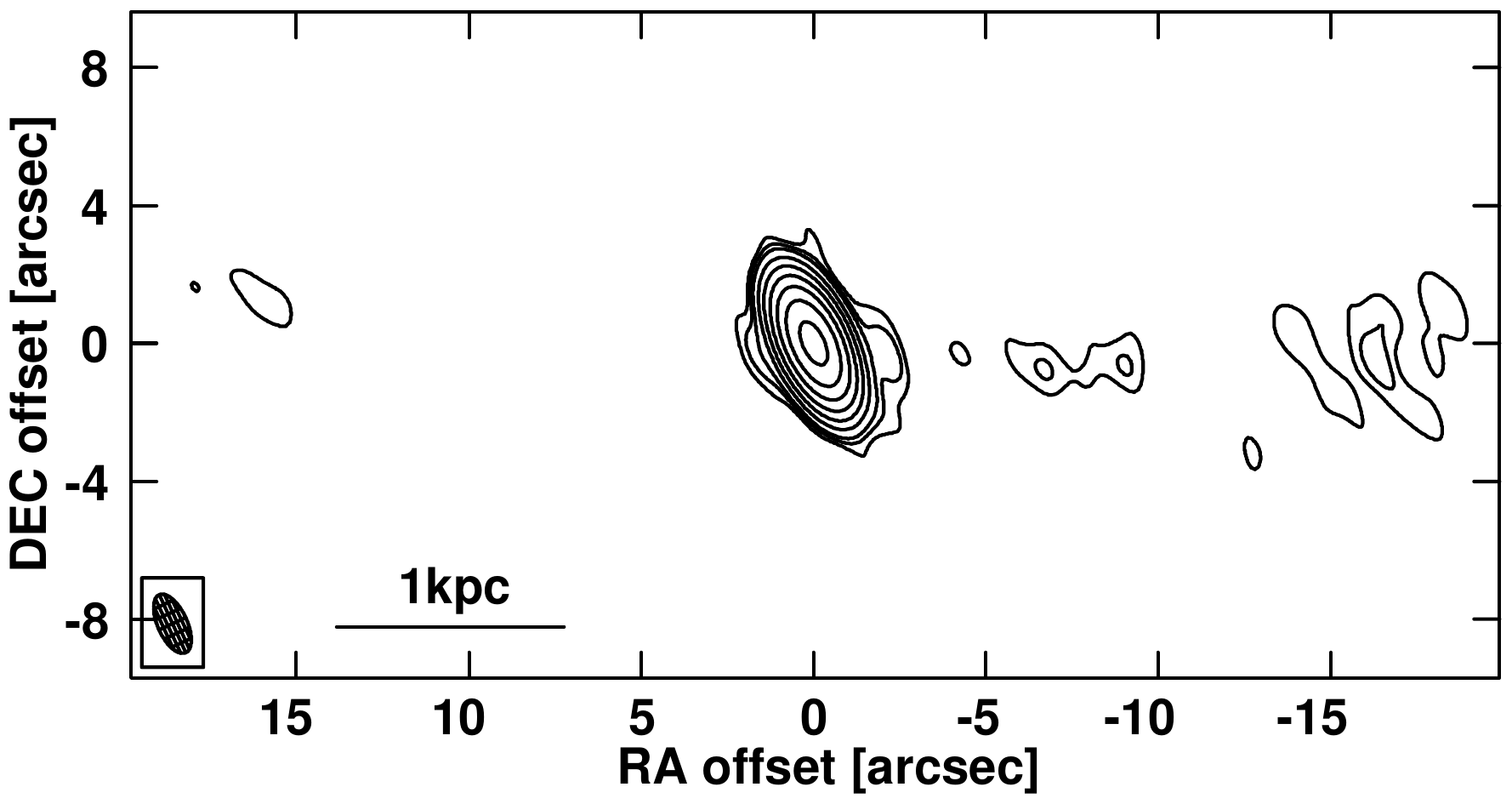}}

\caption{Continuum map of NGC~4261 at 80~GHz. 
Contours begin at the $3\sigma$ level and 
increase by factors of $\sqrt{3}$ to 
$9\sigma$ and by factors of 3 thereafter, 
where $\sigma=3.8\times10^{-2}$ mJy beam$^{-1}$. 
The position offsets are measured from 
RA(J2000)=$12^{\rm h}19^{\rm m}23^{\rm s}.220$ and 
Dec(J2000)=05$^{\circ}$49$^{\prime}$30$^{\prime\prime}$.775. 
The peak intensity, located at the centre, 
is 0.36 Jy beam$^{-1}$. 
The synthesised beam is 
$1^{\prime\prime}.8\times0^{\prime\prime}.85$ 
at a PA of $25^{\circ}$, 
as indicated by the cross-hatched ellipse 
in the bottom-left corner.
}
    \label{fig:n4261cnt}
    \end{center}
    \end{figure}

Figure~\ref{fig:n4261cnt} shows a bright compact source 
at the phase centre 
and weak jet features spanning 5 kpc 
aligned along the east-west direction. 
The position angle (PA) of the alignment is estimated 
as $87\pm1^{\circ}$ 
via a linear-regression fit to the jet features. 
This estimation is in agreement 
with the 5 GHz radio jet PA of $88\pm2^{\circ}$
imaged with the VLA   
\citep{birkinshaw85}.
The bright compact source is partially resolved into 
a core 
at $\Delta$RA = $0^{\prime\prime}.0$ 
along with the east and west nuclear jet components
at $1^{\prime\prime}.5$ 
and $-2^{\prime\prime}$.0, respectively. 
The peak position of the continuum emission 
coincides with that of the phase centre. 
The flux density of the compact core 
within the central $\pm3^{\prime\prime}$
is measured as $S_{\rm 80GHz}$ = 360 mJy 
by means of a two-dimensional Gaussian fit 
using the AIPS task JMFIT. 
Together with literature flux measurements, 
$S_{\rm 115GHz}$ = $326.34\pm0.82$ mJy 
observed with the IRAM 30 m telescope 
\citep{ocana10}, 
 $S_{\rm 236GHz}$ = $253\pm25$ mJy,   
and 
 $S_{\rm 348GHz}$ = $223\pm22$ mJy 
from Atacama Large Millimeter/submillimeter Array projects 
2017.1.00301.S and 2017.1.01638.S 
(Boizelle, priv. communication), 
we find a spectrum $S_\nu \propto \nu^\alpha$ 
with $\alpha = -0.34\pm 0.02$ in the 80-348 GHz range. 
This core spectrum is very inconsistent with thermal emission, 
although it is still somewhat shallower than 
canonical synchrotron values measured in the extended jets, 
suggesting a partially optically thick core environment.

   \begin{figure*}[!ht]
   \begin{center}
   \includegraphics[width=17cm,bb=0 0 504 216]{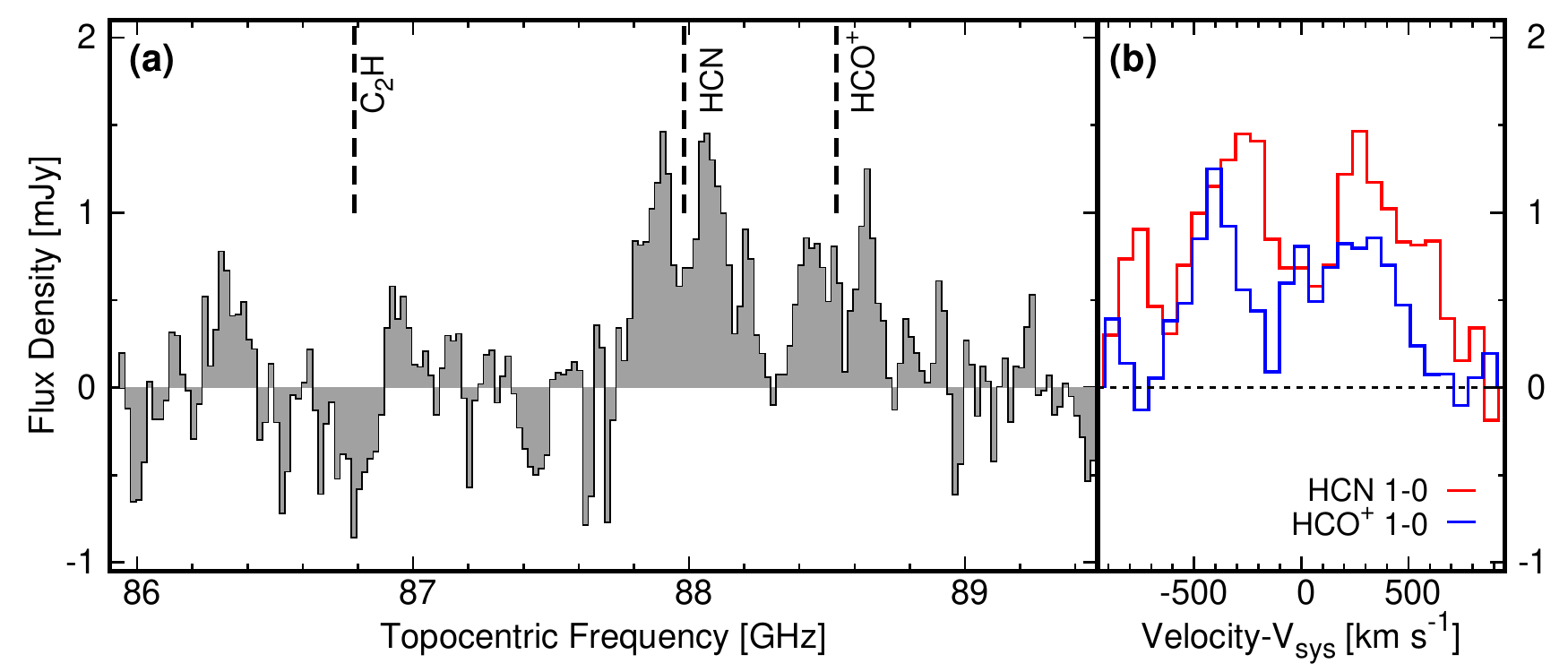}
   \end{center}
\caption{ 
Integrated spectrum over the region within $\pm1^{\prime\prime}$ 
of the core from 85.886 to 89.946 GHz, 
after subtracting continuum emission. 
The frequency resolution is 20~MHz, 
and the rms noise is 0.28 mJy. 
(a) Spectrum as functions of topocentric frequency in GHz.
The vertical dashed lines represent $V_{\rm sys}$ 
for each molecular line.
(b) 
Spectra of HCN $J$=1--0 and HCO$^+$ $J$=1--0 emission lines 
as functions of velocity with respect to $V_{\rm sys}$.
}
\label{fig:n4261spc}
\end{figure*}

\begin{figure*}[!ht]
\centering
\includegraphics[clip,width=15cm]{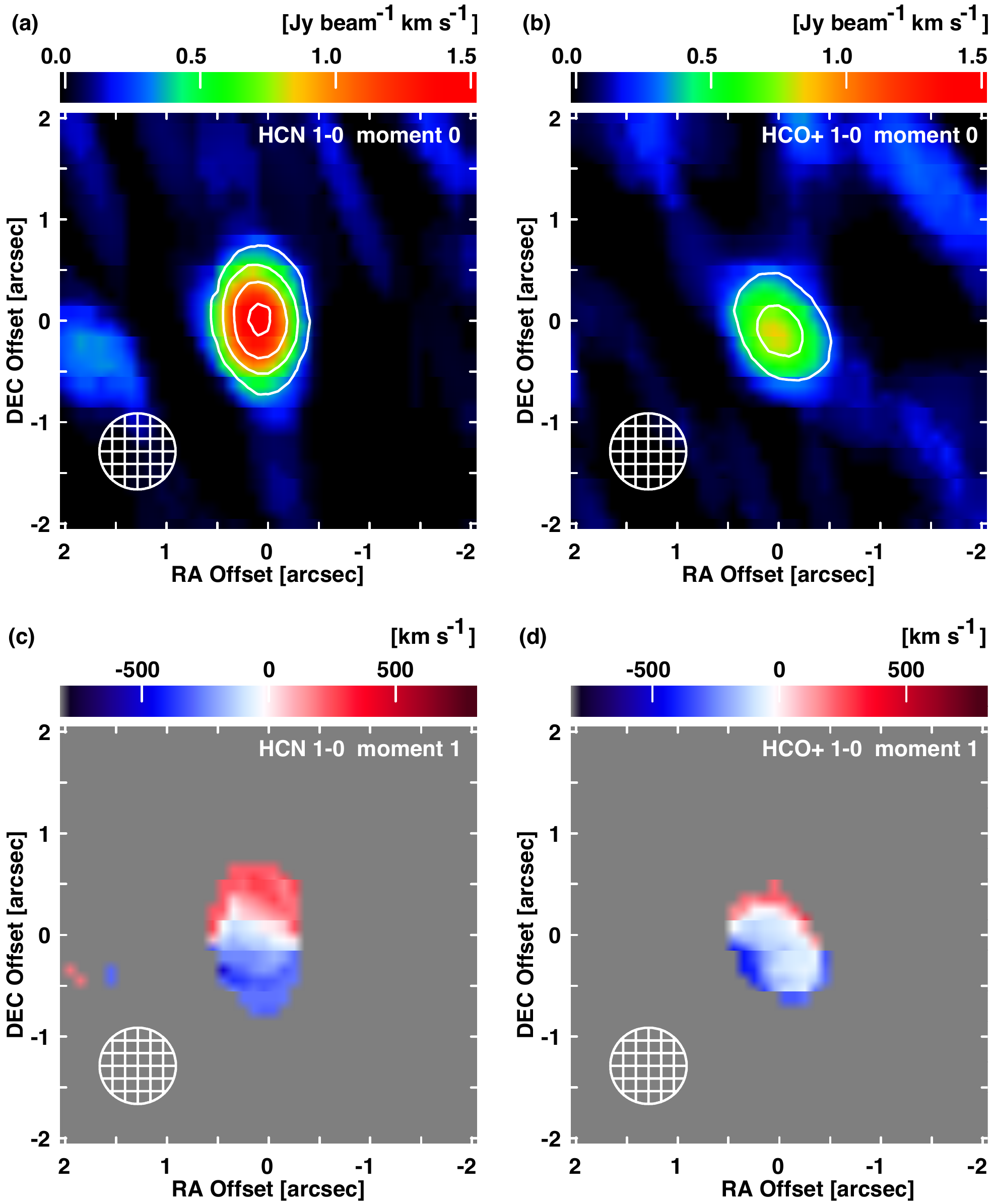}
\caption{
Integrated intensity (moment-0) maps of 
(a) HCN $J$=1--0 and (b) HCO$^{+}$ $J$=1--0 lines 
towards the central $\pm2^{\prime\prime}$.
Both moment-0 images use the same colour scaling 
from 0 to 1.50 Jy beam$^{-1}$ km s$^{-1}$ 
and the same contour levels in steps of $4\sigma$,  
where $\sigma=8.8\times10^{-2}$ Jy beam$^{-1}$ km s$^{-1}$.
The peak intensity is 1.50 and 0.87 Jy beam$^{-1}$ km s$^{-1}$ 
for HCN and HCO$^{+}$, respectively. 
Intensity-weighted velocity (moment-1) maps of 
(c) HCN $J$=1--0 and (d) HCO$^{+}$ $J$=1--0 lines 
towards the central $\pm2^{\prime\prime}$, 
with the same color scaling 
from $-800$ to $+800$ km s$^{-1}$ 
in relative velocity with respect to $V_{sys}$. 
All maps are restored with a circular Gaussian 
beam of $0^{\prime\prime}.75$, 
which is the typical minor-axis size of the synthesised beam.
The position offsets are measured from 
RA(J2000)=$12^{\rm h}19^{\rm m}23^{\rm s}.220$ and 
Dec(J2000)=05$^{\circ}$49$^{\prime}$30$^{\prime\prime}$.775. 
}
\label{fig:moment}
\end{figure*}


\subsection{HCN and HCO$^{+}$ emission}

The spectral profiles of 
the HCN $J$=1--0 and HCO$^+$ $J$=1--0 emission lines 
integrated over the region 
within the central $\pm1^{\prime\prime}$ 
are shown in Figure~\ref{fig:n4261spc}. 
Both molecular lines are detected 
above the $3\sigma$ level 
within a velocity range of $\pm700$ km~s$^{-1}$ 
from $V_{\rm sys}$.
Furthermore, both lines are 
below the $3\sigma$ level 
in the channels around $V_{\rm sys}$ 
and 
exhibit a nearly symmetrical double-peaked spectral profile. 
The double-peaked spectra resemble the double-horned profile 
expected from an inclined rotating disk 
with a central depression or a cavity
\citep[e.g.][]{wiklind97}. 
In addition, a possible absorption feature is detected 
at the redshifted frequencies of C$_2$H $N$=1--0 
with a significance of $3\sigma$ 
in the deepest absorption channel. 
The detected feature could be a blend of six hyperfine components of C$_2$H $N$=1--0 
($J$=3/2--1/2, $F$=1--1, 2--1, 1--0, and $J$=1/2--1/2, $F$=1--1, 0--1, 1--0).

The velocity-integrated intensity (moment-0) maps  
of the HCN $J$=1--0 and HCO$^+$ $J$=1--0 emission lines 
shown in Figures~\ref{fig:moment}(a) and (b)    
reveal a single component, 
which spatially coincides with the central continuum peak. 
A faint feature can be seen $1^{\prime\prime}.5$ east 
of the phase centre in the HCN moment-0 map, 
but it does not reach the $4\sigma$ level. 
A least-squares ellipse fit 
to the regions defined by the 
$4\sigma$ contour of the integrated intensity 
for each moment-0 map 
is listed in Table~\ref{tab:ellifit}. 
The extent of the significant HCN emission 
spans $1^{\prime\prime}.4$ (210 pc) 
along the north-south direction (PA$=2^{\circ}$), 
which is 
slightly longer than 
a beam size of $0^{\prime\prime}.75$. 
The distribution of HCO$^{+}$ emission is more concentrated 
at the centre 
and the HCO$^+$ component is fainter. 
Both of these molecular lines 
originate from the same $1^{\prime\prime}.7$-diameter  dust disk 
found in the HST images \citep{jaffe93,jaffe96} 
and are more centrally concentrated 
compared to the CO $J$=2--1 emission 
that spans $2^{\prime\prime}$ 
in prior interferometric observations \citep{boizelle21}. 
The intensity-weighted velocity (moment-1) map of HCN 
(Figure~\ref{fig:moment}(c))
tentatively 
shows a velocity gradient 
along the major axis, 
perpendicular to the jet PA. 
The distribution and velocity structure of the HCN line 
are in agreement with those obtained for  
CO $J$=2--1 and $J$=3--2 lines 
\citep{boizelle21}. 
Furthermore, 
the moment-1 map of HCO$^+$ (Figure~\ref{fig:moment}(d))
roughly follows  
the velocity gradient along the north-south direction, 
although the velocity gradient is less evident 
than that of HCN. 
It should be noted that 
the HCO$^+$ distribution appears to exhibit 
a barely resolved disk structure. 
Thus, multiple velocity features should be spatially unresolved. 

The HCN $J$=1--0 to HCO$^{+}$ $J$=1--0 ratio ($R_{\rm HCN/HCO^{+}}$)  
and the HCN $J$=1--0 to CO $J$=1--0 ratio ($R_{\rm HCN/CO}$) 
are proposed to be good indicators of an AGN-dominated environment in SGs 
\citep{kohno01, kohno05}.  
Velocity-integrated flux densities of 
HCN ($S_{\rm HCN} \Delta V$)
and HCO$^+$ ($S_{\rm HCO^{+}} \Delta V$)
within the central $\pm1^{\prime\prime}$ are  
1.48 and 0.79 Jy km s$^{-1}$, respectively. 
We derive the $R_{\rm HCN/HCO^{+}}$ 
= 1.87 
on sub-kiloparsec scales, 
which is consistent with the mean ratio of $1.84\pm0.43$ 
for a sample of AGN host galaxies \citep{privon15}.  
Assuming the CO $J$=2--1 to $J$=1--0 intensity ratio $R_{\rm 21}$ = 0.79 
from the xCOLD (extended CO Legacy Database) for GASS (GALEX Arecibo SDSS Survey) 
sample of nearby galaxies \citep{koss21} 
and the velocity-integrated flux density of CO $J$=2--1 of 3.06 Jy km s$^{-1}$ 
measured for NGC~4261 \citep{boizelle21}, 
we get $R_{\rm HCN/CO}$ = 0.38. 
The resultant line ratios $R_{\rm HCN/HCO^{+}}$ of 1.87 and 
$R_{\rm HCN/CO}$ of 0.38 in NGC~4261 are typical values 
expected for `pure' AGNs 
with the absence of any associated nuclear starburst 
activity \citep{kohno05}.

   \begin{table}
      \caption{Elliptical fit parameters to integrated intensities of HCN and HCO$^+$}
	\label{tab:ellifit}
	\begin{tabular}{c c c c c}
            \hline\hline
	Line &
	$\Delta{\rm RA}$ & 
	$\Delta{\rm Dec}$ & 
	$\theta_{\rm mj}\times\theta_{\rm mn}$ &
	PA \\
 & [$^{\prime\prime}$] 
 & [$^{\prime\prime}$] 
 & [$^{\prime\prime}\times^{\prime\prime}$]
 & [$^\circ$] \\  
 (1) & (2) & (3) & (4) & (5) \\          
            \hline
HCN(1--0)     
    & $0.08\pm0.01$ 
    & $0.02\pm0.01$ 
    & $1.4\times0.8$
    & 2 \\
HCO$^+$(1--0) 
    & $-0.04\pm0.02$ 
    & $-0.07\pm0.02$
    & $1.0\times0.7$
    & 35 \\
            \hline
\end{tabular}
\tablefoot{
Column(1): Line species. 
(2): RA offset from the continuum source centroid 
at RA(J2000)=$12^{\rm h}19^{\rm m}23^{\rm s}.220$. 
(3): DEC offset from the continuum source centroid 
at 
DEC(J2000)=05$^{\circ}$49$^{\prime}$30$^{\prime\prime}$.775.
(4): Angular widths of the major and minor axes.
(5): Position angle. 
}
   \end{table}

\section{Discussions}

\begin{figure}[t]
\begin{center}
\includegraphics[width=7cm,bb=0 0 214 289]{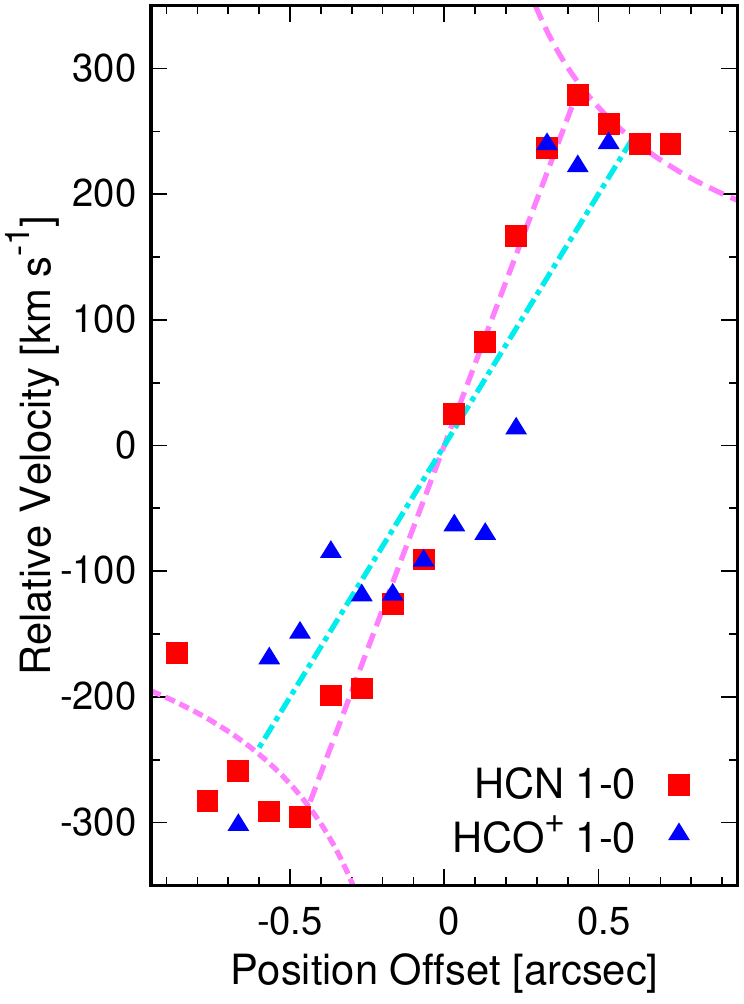}
\caption{
Luminosity-weighted moment-1 measurements in declination for HCN and HCO$^+$ lines.
The dashed magenta line represents 
the best-fit Keplerian rotation model 
to the HCN data (red squares).
The dot-dashed cyan line indicates 
a linear gradient fit to the HCO$^+$ data (blue triangles). 
The straight dashed lines in magenta and cyan 
correspond to a rotating gas ring with a radius of 66 and 92 pc, respectively. 
}
\label{fig:pv}
\end{center}
\end{figure}

\subsection{Keplerian rotation of CNDs}

Luminosity-weighted moment-1 measurements 
for $J$=1--0 emission lines of HCN and HCO$^+$  
along the major axis (PA$=0^{\circ}$) 
are shown in Figure~\ref{fig:pv}. 
Data points are derived from the velocity slice 
across the centre along PA$=0^{\circ}$ 
in the HCN and HCO$^+$ moment-1 maps 
(Figure~\ref{fig:moment}(c)(d)).  
We performed a linear fit to the HCN data points 
at a position offset within $\pm0^{\prime\prime}.4$
and a Keplerian rotation fit to the HCN data 
at $<-0^{\prime\prime}.4$ and $>+0^{\prime\prime}.4$.  
The HCN data points and 
their best-fit Keplerian rotation curves 
indicate that 
the HCN emission traces the rotation with 
a radius in the range 66--130 pc. 
The enclosed mass estimated from the Keplerian rotation fitting is 
$(1.6\pm0.1)\times10^9$ $M_{\odot}$, 
after adopting a disk inclination angle of $64^{\circ}$
\citep{ferrarese96}. 
This is in good agreement with the black hole mass measurement 
made using the CO lines  
($1.67\times10^9 M_{\odot}$; \citealt{boizelle21}), 
while it is three times larger  
than the mass determination inside $0^{\prime\prime}.1$ 
(14.5 pc) 
based on the ionised gas kinematics 
($4.9\times10^8 M_{\odot}$; \citealt{ferrarese96}).
The HCO$^+$ data also appear to display a velocity gradient 
along the major axis, 
while the data points are more scattered 
from the linear gradient. 
This could be due to 
the barely resolved HCO$^+$ multiple velocity features. 
The best-fit rotational gas model 
for the HCO$^+$ data is 
consistent with the Keplerian rotational model 
obtained from the HCN data. 
This implies that 
HCN and HCO$^+$ emission trace the same galaxy potential 
for the same radii of approximately 60--130 pc.

\subsection{Mass of dense molecular gas in CNDs}

The HCN line luminosity 
has been used to estimate the mass of the dense molecular gas, 
by applying the HCN luminosity-to-mass conversion factor 
$\alpha_{\rm HCN}$. 

Using the HCN line luminosity 
in accordance with \cite{solomon97} and \cite{tan18}, 
the dense molecular gas mass ($M_{\rm dg}$) is as follows:  
\begin{eqnarray}
M_{\rm dg} &=& \alpha_{\rm HCN} L_{\rm HCN}^{\prime} \nonumber \\
    &=& 3.25\times10^7 \alpha_{\rm HCN} S_{\rm HCN}\Delta V 
\nu_{\rm obs}^{-2} D_{\rm L}^2 (1+z)^{-3} 
 ~ M_{\odot},
\end{eqnarray}
where 
$\alpha_{\rm HCN}$ denotes the HCN luminosity-to-dense-gas-mass
conversion factor in $M_{\odot}$ (K km s$^{-1}$ pc$^{2}$)$^{-1}$, 
$L_{\rm HCN}^{\prime}$ denotes the HCN luminosity 
in K~km~s$^{-1}$~pc$^2$, 
$S_{\rm HCN} \Delta V$ denotes the velocity-integrated flux density 
of HCN in Jy~km~s$^{-1}$, 
$\nu_{\rm obs}$ denotes the observed line frequency in GHz 
and $D_{\rm L}$ denotes the luminosity distance in Mpc. 
We obtain $L_{\rm HCN}^{\prime}$ = $6.03\times10^6$ K~km~s$^{-1}$~pc$^2$.

The luminosity-to-mass conversion factor, $\alpha_{\rm HCN}$,  
can vary from 
0.24 to over 20 $M_{\odot}$ (K km~s$^{-1}$ pc$^2$)$^{-1}$ 
\citep{barcos18,evans20,jones21}, 
depending on the gas density and the line opacity  
\citep{jones21,wang21}. 
If we consider the standard extragalactic conversion factor 
$\alpha_{\rm HCN}$ of  
10 $M_{\odot}$ (K km~s$^{-1}$ pc$^2$)$^{-1}$ 
\citep{gao04}, 
$M_{\rm dg}$ is calculated to be   
$6.03\times10^7$  $M_{\odot}$. 
This value is over five times higher than the total gas mass 
$M_{\rm gas}$ = $1.12\times10^7$  $M_{\odot}$ 
reported by \citet{boizelle21} 
using a typical CO conversion. 
The adopted conversion factor $\alpha_{\rm HCN}$ = 10 could be 
an order of magnitude too large 
because NGC~4261 has a higher $R_{\rm HCN/CO}$ compared with other galaxies. 
The values of $M_{\rm dg}$ and $M_{\rm gas}$ ($10^{7}$ $M_{\odot}$)  
 are consistent with CND masses on 100 pc scales 
 in galaxies of various types, including SGs, ETGs, and RGs 
\citep[e.g.][]{izumi16,boizelle17,garcia19,ruffa19}. 
In contrast, 
it is significantly more massive than 
the molecular gas mass of the 100 pc CND 
in other nearby RGs:  
NGC~5128 ($2 \times 10^6$ $M_{\odot}$; 
\citealt{mccoy17})
and NGC~1052 
($5.3 \times 10^5$ $M_{\odot}$; \citealt{kameno20}).  
Interestingly, these three RGs  
show dissimilar CND characteristics to one another 
despite many resemblances,  
such as 
the classification of a LINER AGN,
the bright and  two-sided radio jets,  
the presence of a surrounding torus, 
the central condensation of the ionised gas,  
and 
the central high column density of hydrogen 
\citep[e.g.][]{marconi00,kameno01,markowitz07,balokovic21}. 
Further observations in a larger sample of RGs 
are required 
to understand the variety of the observed CND characteristics.

\subsection
{$M_{\rm dg}$--$\dot M_{\rm BH}$ correlation}

\citet{izumi16} report  
a positive correlation between $M_{\rm dg}$ and 
the black hole mass accretion rate, $\dot M_{\rm BH}$,  
for SGs. 
By applying an $M_{\rm dg}$ of $6.03\times10^7$ $M_{\odot}$ 
and 
using our measurement 
to the regression line for $M_{\rm dg}$ and $\dot M_{\rm BH}$ 
offered by \citet{izumi16}, 
the inferred $\dot M_{\rm BH}$ corresponds to 
$10^{-2.48}$ $M_\odot$ yr$^{-1}$.  
This value is comparable to 
an $\dot M_{\rm BH}$ of $10^{-2.70}$ $M_\odot$ yr$^{-1}$, 
 which is obtained by using the 
 $L_{\rm bol}$--$\dot M_{\rm BH}$ relation \citep{alexander12}: 
\begin{equation}
    \dot M_{\rm BH} = 0.15 
        \biggl( \frac{0.1}{\eta} \biggr)
        \biggl( \frac{L_{\rm bol}}{10^{45} {\rm erg~s}^{-1}} \biggr)
 ~ M_\odot {\rm yr}^{-1}, 
\end{equation}
where $\eta$ = 0.1 is a typical value 
for mass--energy efficiency conversion
\citep{marconi04} 
and $L_{\rm bol}$ is equal to $10^{42.6}$ erg s$^{-1}$ for NGC~4261,  
as reported by \citet{hermosa22}.

The derived $M_{\rm dg}$ appears to be in agreement with 
the positive correlation between $M_{\rm dg}$ and 
$\dot M_{\rm BH}$ at the CND scale in NGC~4261. 
It should be noted, however, that there are significant uncertainties 
in $\alpha_{\rm HCN}$, 
 the $L_{\rm 2-10keV}$--$L_{\rm bol}$ relation \citep[e.g.][]{eracleous10}, 
 and 
 the confidence interval in the correct $\eta$ to use 
 in general SGs.

%

\section{Conclusions}

We mapped 
the central 5 kpc  of NGC~4261 with NOEMA 
in the HCN and HCO$^+$ $J$=1--0 lines and  
the 80 GHz continuum. 
The continuum image reveals a core-dominant 
synchrotron jet structure,  
which consists of 
a bright central source and weak jet features 
aligned along the east-west direction. 
HCN and HCO$^{+}$ emission lines are detected in NGC~4261 
for the first time, 
covering a velocity range of 
$\pm700$ km~s$^{-1}$ relative to $V_{\rm sys}$.
The molecular gas is distributed in 
a rotating sub-kiloparsec disk structure, 
which coincides with the bright central source 
in position. 
The Keplerian rotation model obtained from 
the velocity fields of HCN and HCO$^+$  
yields an enclosed mass of $(1.6\pm0.1)\times10^{9}$ 
$M_\odot$. 
Using the HCN line luminosity 
and the standard extragalactic luminosity-to-mass 
conversion factor, 
the dense gas mass, $M_{\rm dg}$,  
associated with the CND is estimated  
to be $6.03\times10^7$  $M_{\odot}$.
This value is comparable to a typical CND mass 
measured in galaxies of various types, 
including SGs, ETGs, and RGs. 
The derived $M_{\rm dg}$ and $\dot M_{\rm BH}$ in NGC~4261 
align with the positive correlation 
between $M_{\rm dg}$ and $\dot M_{\rm BH}$
seen in SGs, 
which supports the scenario that 
star formation in CNDs drives mass accretion onto SMBHs, 
although there are significant uncertainties in the parameters 
of the correlation.

\begin{acknowledgements}

We acknowledge the anonymous referee for valuable comments that improved our manuscript. 
This work is based on IRAM/NOEMA observations carried out under project number W18CK.
IRAM is supported by INSU/CNRS (France), MPG (Germany) and IGN (Spain).
S.S.-S. is supported by JSPS KAKENHI grant No. 21K03628. 
S.K. is supported by JSPS KAKENHI grant No. 18K0371. 

\end{acknowledgements}

%
   \bibliographystyle{aa} 
   \bibliography{aa202244047} 
%
\end{document}